\documentclass[10pt]{article}
 \begin{document}

\title{Multiverses and Cosmology: Philosophical Issues}
\author{ W. R. Stoeger$^{1,2}$, G. F. R. Ellis$^1$, and U.
Kirchner$^1$. }

\maketitle

$^1$ Department of Mathematics and Applied Mathematics,  University of Cape
Town, 7700 Rondebosch, South Africa. \\

$^2$ Permanent Address: Vatican Observatory Research Group, Steward
Observatory, The University of Arizona, Tucson,
Arizona 85721, USA. \\

 \begin{abstract}
The idea of a multiverse -- an ensemble of universes or universe
domains -- has received increasing attention in cosmology, both as
the outcome of the originating process that generated our own
universe, and as an explanation for why our universe appears to be
fine-tuned for life and consciousness. Here we review how
multiverses should be defined, stressing the distinction between
the collection of all possible universes and ensembles of really
existing universes, which distinction is essential for anthropic
arguments. We show that such realised multiverses are by no means
unique, and in general require the existence of a well-defined and
physically motivated distribution function on the space of all
possible universes. Furthermore, a proper measure on these spaces
is also needed, so that probabilities can be calculated. We then
discuss several other physical and philosophical problems arising
in the context of ensembles of universes, including realized
infinities and the issue of fine-tuning -- whether very special or
generic primordial conditions are more fundamental in cosmology.
 Then we briefly summarise scenarios like chaotic inflation, which
suggest how ensembles of universe domains may be generated, and
point out that the regularities underlying any systematic
description of truly disjoint multiverses must imply some kind of
common generating mechanism, whose testability is problematic.
Finally, we discuss the
issue of testability, which underlies the question of whether
multiverse proposals are really scientific propositions rather
than metaphysical proposals.
 \end{abstract}

 \section{Introduction}
  Over the past twenty years the proposal of a really existing ensemble of
universes -- a `multiverse' -- has gained prominence in cosmology,
even though there is so far only inadequate theoretical and
observational support for its existence. The popularity of this
proposal can be traced to two factors. The first is that quite a
few promising programs of research in quantum and very early
 universe cosmology suggest that the very processes which could have brought our
  universe or region of the universe into existence from a primordial quantum
  configuration, would have generated many other universes or universe regions
as well.  This was first modelled in a specific way by Vilenkin
(1983) and was developed by Linde (Linde 1983, 1990) in his
chaotic cosmology scenario. Since then many others, e. g. Leslie
(1996), Weinberg (2000), Sciama (1993), Deutsch (1998), Tegmark
(1998, 2003), Smolin (1999), Lewis (2000),  Weinberg (2000), and
Rees (2001) have discussed ways in which an ensemble of universes
or universe domains might originate physically. More recently
specific impetus has been given to this possibility by superstring
theory. It is now claimed by some that versions of these theories
provide ``landscapes'' populated by a large number of vacua, * each
of which could occur in or initiate a separate universe domain,* with different
values of the physical parameters, such as the cosmological
constant, the masses of the elementary particles and the strengths
of their interactions  (Kachru, et al. 2003; Susskind 2003, 2005,
and references therein).

So far, none of these proposals has been developed to the point of
actually describing such ensembles of universes in detail, nor has
it been demonstrated that a generic well-defined ensemble will
admit life. Some writers tend to imply that there is only one
possible multiverse, characterised by ``all that can exist does
exist'' (Lewis 2000, see also Gardner 2003). This vague
prescription actually allows a vast variety of different
realisations with differing properties, leading to major problems
in the definition of the ensembles and in averaging, due to the
lack of a well-defined measure and the infinite character of the
ensemble itself. Furthermore, it is not at all clear that we shall
ever be able to accurately delineate the class of all possible
universes.

  The second factor stimulating the popularity of multiverses is that it is
the only scientific way of avoiding the fine-tuning seemingly
required for our universe. This applies firstly to the
cosmological constant, which seems to be fine-tuned by 120 orders
of magnitude relative to what is expected on the basis of quantum
field theory (Weinberg 2000, Susskind 2005). If (almost) all
values of the cosmological constant occur in a multiverse, then we
can plausibly live in one with the very low observed non-zero
value; indeed such a low value is required in order that galaxies,
stars, and planets exist and provide us with a suitable habitat
for life.

This is an example of the second motivation, namely the `anthropic
principle' connection: If any of a large number of parameters
which characterize our universe -- including fundamental
constants, the cosmological constant, and and initial conditions
-- were slightly different, our universe would not be suitable for
complexity or life. What explains the precise adjustment of these
parameters so that microscopic and macroscopic complexity and life
eventually emerged? One can introduce a ``Creator" who
intentionally sets their values to assure the eventual development
of complexity. But this move takes us beyond science. The
existence of a large collection of universes, which represents the
full range of possible parameter values, though not providing an
ultimate explanation, would provide a scientifically accessible
way of avoiding the need for such fine-tuning. If physical
cosmogonic processes naturally produced  such a variety of
universes, one of which was ours, then the puzzle of fine- tuning
is solved. We simply find ourselves in one in which all the many
conditions for life have been fulfilled.

 Of course, through cosmology we must then discover and describe the
 process by which that collection of diverse universes, or universe
 domains, was generated, or at least could have been generated, with
 the full range of characteristics they possess. This
  may be possible. It is analogous to the way in which we look upon the
  special character of our Solar System. We do not agonize how initial
conditions for the Earth and Sun were specially set so that life
would eventually emerge -- though at some level that is still a
mystery. We simply realize that our Solar System in one of
hundreds of billions of others in the Milky Way, and accept that,
though the probability that any one of them is bio-friendly is
very low, at least a few of them will naturally be so. We have
emerged as observers in one of those. No direct fine-tuning is
required, provided we take for granted both the nature of the laws
of physics and the specific initial conditions in the universe.
The physical processes of stellar formation throughout our galaxy
naturally leads to the generation of the full range of possible
stellar systems and planets.

  Before going on, it is necessary to clarify our terminology.
Some refer to the separate expanding universe regions in chaotic
inflation as `universes', even though they have a common causal
origin and are all part of the same single spacetime. In our view
(as `uni' means `one') \textit{the Universe} is by definition the
one unique connected\footnote{ \textquotedblleft
Connected\textquotedblright\ implies \textquotedblleft Locally
causally connected\textquotedblright , that is all universe
domains are connected by $C^{0}$ timelike lines which allow any
number of reversals in their direction of time, as in Feynman's
approach to electrodynamics. Thus, it is a union of regions that
are causally connected to each other, and transcends particle and
event horizons; for example all points in de Sitter space time are
connected to each other by such lines.} existing spacetime of
which our observed expanding cosmological domain is a part. We
will refer to situations such as in chaotic inflation as a
\textit{Multi-Domain Universe}, as opposed to a completely
causally disconnected \textit{Multiverse}. Throughout this paper,
when our discussion pertains equally well to disjoint collections
of universes (multiverses in the strict sense) and to the
different domains of a Multi-Domain Universe, we shall for
simplicity simply use the word \textquotedblleft
\textit{ensemble}\textquotedblright . When the universes of an
ensemble are all sub-regions of a larger connected spacetime - the
\textquotedblleft Universe as a whole\textquotedblright - we have
the multi-domain situation, which should be described as such.
Then we can reserve \textquotedblleft
multiverse\textquotedblright\ for the collection of genuinely
disconnected \textquotedblleft universes\textquotedblright\ --
those which are not locally causally related.

  In this article, we shall critically examine the concept of an ensemble of
universes or universe domains, from both physical and
philosophical points of view, reviewing how they are to be defined
physically and mathematically in cosmology (Ellis, Kirchner and
Stoeger 2003, hereafter referred to as EKS), how their existence
could conceivably be validated scientifically, and focusing upon
some of the key philosophical problems associated with them.
  We have already addressed the physics and cosmology of such ensembles in a
previous paper (EKS), along with some limited discussion
  of philosophical issues. Here we shall summarize the principal conclusions of
that paper and then discuss in detail the more philosophical
issues.

  First of all, we review the description of the the set of possible universes and sets of
  realised (i. e., really existing) universes and the relationship between these
two kinds of sets. It is fundamental to have a general
provisionally adequate scheme to describe the set of all possible
universes. Using this we can then move forward to describe
potential sets of actually existing universes by defining
distribution functions (discrete or continuous) on the space of
possible universes. A given distribution function indicates which
of the theoretically possible universes have been actualized to
give us a really existing ensemble of universes or universe
domains. It is obviously crucial to maintain the distinction
between the set of all possible universes, and the set of all
existing universes. For it is the set of all existing universes
which needs to be explained by cosmology and physics -- that is,
by a primordial originating process or processes. Furthermore, it
is only an {\it actually existing} ensemble of universes with the
required range of properties which can provide an explanation for
the existence of our bio-friendly universe without fine-tuning
(see also McMullin 1993, p. 371). A conceptually possible ensemble
is not sufficient for this purpose -- one needs universes which
actually exist, along with mechanisms which generate their
existence. We consider in some depth how the existence of such an
actually existing ensemble might be probed experimentally and
observationally - this is the key issue determining whether the
proposal is truly a scientific one or not.

Though the ensemble of all possible universes is undoubtedly
infinite, having an infinite ensemble of actually existing
universes is problematic -- and furthermore blocks our ability to
assign statistical measures to it, as we shall discuss in some
detail later. For all these reasons, any adequate cosmological
account of the origin of our universe as one of a collection of
many universes -- or even as a single realised universe -- must
include a process whereby the realised ensemble is selected from
the space of all possible universes and physically generated.
 But it must also provide some metaphysical view on the origin of the set of
possible universes as a subset of the set of conceivable universes
- which is itself a very difficult set to define\footnote{Science
fiction and fantasy provide a rich treasury of conceivable
universes, many of which will not be "possible universes" as
outlined above.}.

\section{Describing Ensembles: Possibility}

To characterize an ensemble of existing universes, we first need to develop
adequate methods for describing the class of all possible universes. This itself
is philosophically controversial, as it depends very much on what we regard as
"possible." At the very least, describing the class of all possible universes
requires us to specify, at least in principle, all the ways in which
universes can be different from one another, in terms of their physics,
chemistry, biology, etc. We have done this in EKS, which we shall review here.

\subsection{The Set of Possible Universes}

Ensembles of universes, or multiverses, are most easily represented classically by the
structure and the dynamics of a space $\mathcal{M}$ of all possible
universes, each of which can be described in terms of a set of states $s$
in a state space $\mathcal{S}$ (EKS). Each universe $m$ in $\mathcal{M}$ will be
characterised by a set $\mathcal{P}$ of distinguishing parameters $p$, which
are coordinates on $\mathcal{S}$ (EKS). Each $m$ will evolve
from its initial state to some final state according to the operative dynamics,
 with some or all of its parameters varying as it does so. The
course of this evolution of states will be represented by a path
in the state space $\mathcal{S}$. Thus, each such path (in
degenerate cases, a point) is a representation of one of the
universes $m$ in $\mathcal{M}$. The parameter space $\mathcal{P}$
has dimension $N$ which is the dimension of the space of models
$\mathcal{M}$; the space of states $ \mathcal{S}$ has $N+1$
dimensions, the extra dimension indicating the change of each
model's states with time, characterised by an extra parameter,
e.g., the Hubble parameter $H$ which does not distinguish between
models but rather determines what is the state of dynamical
evolution of each model. Note that $N$ may be infinite, and indeed
will be so unless we consider only geometrically highly restricted
sets of universes.

This classical, non-quantum-cosmological formulation of the set of
all possible universes is obviously provisional and not
fundamental. Much less should it provide the basis for
adjudicating the ontology of these ensembles and their
components.\footnote{We thank and acknowledge the contribution of
an anonymous referee who has pointed this out to us, and has
stimulated this brief discussion of the important role of quantum
cosmology in defining multiverses.}  It provides us with a
preliminary systematic framework, consistent with our present
limited understanding of cosmology,  within which to begin
studying ensembles of universes and universe domains. It is
becoming very clear that, from what we are  beginning to learn
from quantum cosmology, a more fundamental framework will have to
be developed that takes seriously quantum issues such as
entanglement. Additionally, there are serious unresolved problems
concerning time in quantum cosmology. Already at the level of
general relativity itself, as everyone recognizes, time loses its
fundamental, distinct character. What is given is space-time, not
space and time. Time is now intrinsic to a given universe domain
and its dynamics and there is no preferred or unique way of
defining it (Isham 1988, 1993; Smolin 1991; Barbour 1994; Rovelli
2004; and references therein). When we go to quantum gravity and
quantum cosmology, time, while remaining intrinsic, recedes
further in prominence and even seems to disappear. The Wheeler-de
Witt equation for ``wave function of the universe,'' for example,
does not explicitly involve time -- it is a time-independent
equation. However, our provisional classical formulation receives
support from the fact that dynamics and an intrinsic time appear
to emerge from it as the universe expands out of the Planck era
(see, for instance, Isham 1988 and Rovelli 2004, especially pp.
296-301). Furthermore, as yet there is no adequate quantum gravity
theory nor quantum cosmological resolution to this issue of the
origin and the fundamental character of time -- just tantalizing
pieces of a much larger picture. The only viable approach at present is to
proceed on the basis of the emergent classical description.

And then there are related issues connected with decoherence --
how is the transition from ``the wave function of the universe''
to the classical universe, or an ensemble of universe domains,
effected, and what emerges in this transition? What is crucial
here is that as the wave function decoheres an entire ensemble of
universes or universe domains may emerge. These would all be
entangled with one another. This would provide the fundamental
basis for the quantum ontology of the ensemble.\footnote{Again, we
thank the same referee for emphasizing the importance of the
possibility.} Furthermore, it would provide a fundamental
connection among a large number of the members of our classically
defined $\mathcal{M}$ above. We have already stressed the
difference between a multi-domain universe and a true multiverse.
An entangled ensemble of universe domains decohering from a
cosmological wave function would be an important example of that
case. This process of cosmological decoherence, which we as yet do
not understand and have not adequately modelled , may turn out to
be a key generating mechanism for a really existing multiverse. In
that case we would want to define a much more fundamental space of
all possible cosmological wave functions. Each of these would
generate an ensemble of classical universes or universe domains
which we have represented individually in $\mathcal{M}$. We could
then map the wavefunctions in that more fundamental space into the
$m$ of $\mathcal{M}$. As yet, however, we do not have even a
minimally reliable quantum cosmology that would enable us to
implement that.

Despite our lack of understanding at the quantum cosmological
level, and the less than fundamental character of our space
$\mathcal{M}$, it enables us proceed with our discussion of
cosmological ensembles at the non-quantum level - which is what
cosmological observations relate to. While doing so, we must keep
the quantum cosmological perspective in mind. Though we are
without the resources to elaborate it more fully, it provides a
valuable context within which to interpret, evaluate and critique
our more modest classical discussion here.

Returning to our description of the space $\mathcal{M}$ of
possible universes $m$, we must recognize that it is based on an
assumed set of laws of behaviour, either laws of physics or meta-laws
that determine the laws of physics, which all $m$ have in
common. Without this, we have no basis for defining it. Its
overall characterisation must therefore incorporate a description
both of the geometry of the allowed universes and of the physics
of matter. Thus the set of parameters $\mathcal{P}$ will include
both geometric and physical parameters.

Among the important subsets of the space $\mathcal{M}$ are (EKS):
$\mathcal{M}_{\mathrm{FLRW}}$, the subset of all possible
Friedmann-Lema\^{\i}tre-Robertson-Walker (FLRW) universes, which
are exactly isotropic and spatially homogeneous;
$\mathcal{M}_{\mathrm{almost-FLRW}}$, the subset of all universes
which deviate from exact FLRW models by only small, linearly
growing anisotropies and inhomogeneities;
$\mathcal{M}_{\mathrm{anthropic}}$, the subset of all possible
universes in which life emerges at some stage in their evolution.
This subset intersects $\mathcal{M}_{\mathrm{almost-FLRW}}$, and
may even be a subset of $\mathcal{M}_{\mathrm{almost-FLRW}},$ but
does not intersect $ \mathcal{M}_{\mathrm{FLRW}}$, since realistic
models of a life-bearing universe like ours cannot be exactly
FLRW, for then there is no structure.

If $\mathcal{M}$ truly represents all possibilities, as we have
already emphasized, one must have a description that is wide
enough to encompass \textit{all} possibilities. It is here that
major issues arise: how do we decide what all the possibilities
are? What are the limits of possibility? What classifications of
possibility are to be included? From these considerations we have
the first key issue (EKS): \newline

\noindent \textbf{Issue 1:} What determines $\mathcal{M}$? Where
does this structure come from? What is the meta-cause, or ground,
that delimits this set of possibilities? Why is there a uniform
structure across all universes $m$ in $ \mathcal{M}$?
\newline

It should be obvious that these same questions would also have to
be addressed with regard to the more fundamental space of all
cosmological wave functions we briefly described earlier, which
would probably underlie any ensembles of universes or universe
domains drawn from $\mathcal{M}$. It is clear, as we have
discussed in EKS, that these questions cannot be answered
scientifically, though scientific input is necessary for doing so.
How can we answer them philosophically?

\subsection{Adequately Specifying Possible Anthropic Universes}

When defining any ensemble of universes, possible or realised, we must
specify all the parameters which differentiate members of the ensemble from
one another at any time in their evolution. The values of these parameters
may not be known or determinable initially in many cases -- some of them may
only be set by transitions that occur via processes like symmetry breaking
within given members of the ensemble. In particular, some of the parameters
whose values are important for the origination and support of life may only
be fixed later in the evolution of universes in the ensemble.

We can separate our set of parameters $\mathcal{P}$ for the space of all
possible universes $\mathcal{M}$ into different categories, beginning with
the most basic or fundamental, and progressing to more contingent and more
complex categories (see EKS). Ideally they should
 all be independent of one another,
but we will not be able to establish that independence for each parameter,
except for the most fundamental cosmological ones. In order to categorise
our parameters, we can doubly index each parameter $p$ in $\mathcal{P}$ as $
p_j(i)$ such that those for $j=1-2$ describe basic physics, for $j=3-5$
describe the cosmology (given that basic physics), and $j=6-7$ pertain
specifically to emergence of complexity and of life (see EKS for further
details).

Though we did not do so in our first paper EKS, it may be helpful to
add a separate category of parameters $p_8(i)$, which would relate directly
to the emergence of consciousness and self-conscious life, as well as to
the causal effectiveness of self-conscious (human) life -- of ideas,
intentions and goals. It may turn out that all such parameters may be able to
be reduced to those of $p_7(i)$, just as those of $p_6(i)$ and $p_7(i)$ may be
reducible to those of physics. But we also may discover, instead, that such
reducibility is not
possible.

All these parameters will describe the set of
possibilities we are able to characterise on the basis of our accumulated
scientific experience.
This is by no means a statement that ``all that can occur'' is
arbitrary. On the contrary, specifying the set of possible parameters
determines a uniform high-level structure that is obeyed by all universes in
$\mathcal{M}$.

  In the companion cosmology/physics paper to this one (EKS), we develop in
detail
 the geometry, parameters $p_5(i)$, and the physics, parameters $p_1(i)$ to
 $p_4(i)$, of possible universes. There we also examine in detail the FLRW
sector $\mathcal{M}_{FLRW}$ of the ensemble of all possible
universes $\mathcal{M}$ to illustrate the relevant mathematical
and physical issues. We shall not repeat those discussions here,
as they do not directly impact our treatment of the philosophical
issues upon which we are focusing. However, since one of the
primary motivations for developing the multiverse scenario is to
provide a scientific solution to the anthropic fine-tuning
problem, we need to discuss briefly the set of "anthropic"
universes.

\subsection{The Anthropic subset}

The subset of universes that allow intelligent life to emerge is
of particular interest. That means we need a function on the set
of possible universes that describes the probability that life may
evolve. An adaptation of the Drake equation (Drake and Shostak
1998) gives for the expected number of planets with intelligent
life in any particular universe $m$ in an ensemble (EKS),
\begin{equation}
N_{\mathrm{life}}(m)=N_g*N_S*\Pi *F,  \label{life1}
\end{equation}
where $N_g$ is the number of galaxies in the model and $N_S$ the average
number of stars per galaxy. The probability that a star provides a habitat
for life is expressed by the product
\begin{equation}
\Pi =f_S*f_p*n_e  \label{life4}
\end{equation}
and the probability of the emergence of intelligent life,  given such a habitat,
is expressed by the product
\begin{equation}
F=f_l*f_i.  \label{life5}
\end{equation}
Here $f_S$ is the fraction of stars that can provide a suitable environment
for life (they are `Sun-like'), $f_p$ is the fraction of such stars that are
surrounded by planetary systems, $n_e$ is the mean number of planets in each
such system that are suitable habitats for life (they are `Earth-like'), $
f_l $ is the fraction of such planets on which life actually originates, and
$f_i $ represents the fraction of those planets on which there is life where
intelligent beings develop. The anthropic subset of a possibility space is
that set of universes for which $N_{\mathrm{life}}(m)>0.$

The quantities $\{N_g,N_S,f_S,f_p,n_e,f_l,f_i\}$ are functions of the
physical and cosmological parameters characterised above. So there will be
many different representations of this parameter set depending on the degree
to which we try to represent such interrelations.

In EKS, following upon our detailed
treatment of $\mathcal{M}_{FLRW}$ we identify those FLRW universes
 in which the emergence and sustenance
of life is possible on a broad level\footnote{More accurately,
perturbations of these models can allow life -- the exact FLRW
models themselves cannot do so.} -- the necessary cosmological
conditions have been fulfilled allowing existence of galaxies,
stars, and planets if the universe is perturbed, so allowing a
non-zero factor $ N_g*N_S*\Pi $ as discussed above.
 The fraction of these that will actually be life-bearing
depends on the fulfilment of a large number of other conditions represented
by the factor $F=f_l*f_i,\,$ which will also vary across a generic ensemble,
and the above assumes this factor is non-zero.

\section{The Set of Realised Universes}

We have now characterised the set of possible universes. But in any given
existing ensemble, many will not be realised, and some
may be realised many times.
The purpose of this section is to review our formalism (EKS) for specifying
which of
the \textit{possible} universes (characterised above) occur in a particular
\textit{realised} ensemble.

\subsection{A distribution function for realised
universes}

In order to select from $\mathcal{M}$ a set of realised universes
we need to define on $\mathcal{M}$ a distribution function $f(m)$
specifying how many times each type of possible universe $m$ in
$\mathcal{M}$ is realised\footnote{It has been suggested to us
that in mathematical terms it does not make sense to distinguish
identical copies of the same object: they should be identified
with each other because they are essentially the same. But we are
here dealing with physics rather than mathematics, and with real
existence rather than possible existence, and then multiple copies
must be allowed (for example all electrons are identical to each
other; physics would be very different if there were only one
electron in existence).}. The function $f(m)$ expresses the
contingency in any actualisation -- the fact that not every
possible universe has to be realised. Things could have been
different! Thus, $f(m)$ describes the \textit{ensemble of
universes } or \textit{\ multiverse} envisaged as being realised
out of the set of possibilities. In general, these realisations
include only a subset of possible universes, and multiple
realisation of some of them. Even at this early stage of our
discussion we can see that the really existing ensemble of
universes is by no means unique.

From a quantum cosmology perspective we can consider $f(m)$ as given by an
underlying solution of the Wheeler-de Witt equation, by a given superstring
model, or by some other generating mechanism, giving an entangled ensemble
of universes or universe domains.

The class of models considered is determined by all the parameters
which are held constant (`class parameters'). Considering the
varying parameters for a given class (`member parameters'), some
will take only discrete values, but for each one allowed to take
continuous values we need a volume element of the possibility
space $\mathcal{M}$ characterised by parameter increments
$\mathrm{d} p_j(i) $ in all such varying parameters $p_j(i)$. The
volume element will be given by a product
\begin{equation}
\pi =\Pi _{i,j}\,m_{ij}(m)\,\mathrm{d}p_{j}(i)  \label{measure}
\end{equation}
where the product $\Pi _{i,j}$ runs over all continuously varying
member parameters $i,j$ in the possibility space, and the $m_{ij}$
weight the contributions of the different parameter increments
relative to each other. These weights depend on the parameters
$p_{j}(i)$ characterising the universe $m$. The number of
universes corresponding to the set of parameter increments
$\mathrm{d}p_{j}(i)$ will be $\mathrm{d}N$ given by
\begin{equation}
\mathrm{d}N=f(m)\pi  \label{dist1}
\end{equation}
for continuous parameters; for discrete parameters, we add in the
contribution from all allowed parameter values. The total number of universes
in the ensemble will be given by
\begin{equation}
N=\int f(m)\pi  \label{dist2}
\end{equation}
(which will often diverge), where the integral ranges over all allowed
values of the member parameters and we take it to include all relevant
discrete summations. The probable value of any specific quantity $p(m)$
defined on the set of universes will be given by

\begin{equation}
P= \frac{\int p(m)f(m)\pi}{\int f(m)\pi}  \label{prob}
\end{equation}
Such integrals over the space of possibilities give numbers, averages, and
probabilities.

Now it is conceivable that all possibilities are realised -- that all
universes in $\mathcal{M}$ exist at least once. This would mean that the
distribution function
\[
f(m)\neq 0\mathrm{~for~all~}m\in \mathcal{M}.
\]
But there are an infinite number of distribution functions which would
fulfil this condition. So not even a really existing `ensemble of all possible
universes' is unique. In such ensembles, all possible values of each
distinguishing parameter would be represented by its members in all
possible combinations with all other parameters at
least once. One of the problems is that this
means that the integrals associated with such distribution functions
would often diverge, preventing the calculation of probabilities.

From these considerations we have the second key issue: \newline

\noindent \textbf{Issue 2: } What determines $f(m)$? What is the meta-cause
that delimits the set of realisations out of the set of possibilities?
\newline

The answer to this question has to be different from the answer to
\textit{\ Issue 1}, precisely because here we are describing the
contingency of selection of a subset of possibilities for
realisation from the set of all possibilities -- determination of
the latter being what is considered in \textit{Issue 1}. As we saw
in EKS, and as we shall further discuss here (see Section 6),
these questions can, in principle, be partially answered
scientifically. A really existing ensemble of universes or
universe domains demands the operation of a generating process,
which adequately explains the origin of its members with their
ranges of characteristics and their distribution over the
parameters describing them, from a more fundamental potential, a
specific primordial quantum configuration, or the decoherence of a
specific cosmological wave function. That is, there must be a
specific generating process, whatever it is, which determines
$f(m)$. When it comes to the further question, what is responsible
for the operation of this or that specific generating process
rather than some other one which would generate a different
ensemble, we see (EKS) that an adequate answer cannot be given
scientifically. This is the question why the primordial dynamics
leading to the given really existing ensemble of universes is of a
certain type rather than of some other type. Even if we could
establish $f(m)$ in detail, it is difficult to imagine how we
would {\it scientifically} explain why one generating process was
instantiated rather than some other one. The only possibility for
an answer, if any, is via philosophical, or possibly theological,
considerations.

\subsection{Measures and Probabilities}

From what we have seen above, it is clear that $f(m)$ will enable
us to derive numbers and probabilities relative to the realisation
it defines only if we also have determined a unique measure $\pi $
on the ensemble, characterised by a specific choice of the
$\,$weights $m_{ij}(m)$ in (\ref{measure}), where these weights
will depend on the $p_{j}(i)$. There are a number of difficult
challenges we face in doing this, including the lack of a
``natural measure'' on $\mathcal{M}$ in all its coordinates, the
determination of $f(m)$, or its equivalent, from compelling
physical considerations, and the possible divergence of the
probability integrals (see Kirchner and Ellis, 2003). These issues
have been discussed in EKS.

\subsection{The Anthropic subset}

The expression (\ref{life1}) can be used in conjunction with the
distribution function $f(m)$ of universes to determine the expected number of
planets bearing intelligent life arising in the whole ensemble:
\begin{equation}
N_{\mathrm{life}}(E)=\int f(m)*N_g*N_S*f_S*f_p*n_e*f_l*f_i*\pi  \label{life3}
\end{equation}
(which is a particular case of (\ref{prob}) based on (\ref{life1})). An
anthropic ensemble is one for which $N_{\mathrm{life}}(E)>0$. If the
distribution function derives from a probability function, we may combine
the probability functions to get an overall anthropic probability function -
for an example see Weinberg (2000), where it is assumed that
the probability for galaxy formation is the only relevant parameter for the
existence of life. This is equivalent to assuming that $
N_S*f_S*f_p*n_e*f_l*f_i>0$.

This assumption might be acceptable in our physically realised Universe, but
there is no reason to believe it would hold generally in an ensemble because
these parameters will depend on other ensemble parameters, which will vary.

\section{Anthropic Parameters, Complexity and Life}

The astrophysical issues expressed in the product $\Pi$ (the
lower-$j$ parameters: $j \leq 6$) are the easier ones to
investigate anthropically. We can in principle make a cut between
those consistent with the eventual emergence of life and those
incompatible with it by considering each of the factors in $N_g,$
$N_S,$ and $\Pi$ in turn, taking into account their dependence on
the parameters $p_1(i)$ to $p_5(i),$ and only considering the next
factor if all the previous ones are non-zero. In this way we can
assign \textquotedblleft bio-friendly intervals\textquotedblright
\ to the possibility space $\mathcal{M}$. If $\ N_g*N_S*\Pi \,$ is
non-zero we can move on to considering similarly whether $F$ is
non-zero, based on the parameters $p_6(i)$ to $p_8(i)$,
determining if true complexity is possible, which in turn depends
on the physics parameters $p_1(i)$ in a crucial way that is not
fully understood.

As we go to higher-level parameters we will narrow the
number of the number of universes consistent with self-conscious life even more.
Essentially, we shall have the sequence of inequalities:
\[
N_8 < N_7< N_6 < N_5 < N_4 < N_3 < N_2 < N_1,
\]
where $N_j$ is the total number of universes specified by parameters of level
$j$ which are compatible with the eventual emergence of self-conscious life.

This clearly fits very nicely with the Bayesian Inference approach to
probability and provides the beginnings of an implementation of it for these
multiverses. This approach also clearly keeps the distinction between
necessary and sufficient conditions intact throughout. At each level we add
to the necessary conditions for complexity or life, weeding out those
universes which fail to meet any single necessary condition. Sufficiency is
never really reached in our description -- we really do not know the full
set of conditions which achieve sufficiency. Life demands unique
combinations of many different parameter values that must be realised
simultaneously. Higher-order ($j \geq 6$) parameters $p_j(i)$ may not even be
relevant for many universes or universe domains in a given ensemble, since
the structures and processes to which they refer may only be able to
emerge for certain very narrow ranges of the lower-$j$ parameters. It may
also turn out, as we have already mentioned, that higher-level parameters
may be reducible to the lower-level parameters.

It will be impossible at any stage
to characterise that set of $\mathcal{M}$ in which \textit{all} the
conditions \textit{necessary} for the emergence of self-conscious life and
its maintenance have been met, for we do not know what those conditions are
(for example, we do not know if there are forms of life possible that are
not based on carbon and organic chemistry). Nevertheless it is clear that
life demands unique combinations of many different parameter values that
must be realised simultaneously,
  but do not necessarily involve all parameters (for example Hogan \cite{hogan}
suggests that only 8 of the parameters of the standard particle physics model
are involved in the emergence of complexity).
  When we look at these combinations, they
will span a very small subset of the whole parameter space (Davies 2003,
Tegmark 2003).


\section{Problems With Infinity}

When speaking of multiverses or ensembles of universes -- possible or
realised -- the issue of infinity inevitably crops up. Researchers often
envision
an \textit{infinite} set of universes, in which all possibilities are
realised. Can there be an infinite set of really existing universes?
We suggest that the
answer may very well be ``No''. The common perception that this is possible arises from not
 appreciating the precisions in meaning and the restrictions in application
 associated with this profoundly
 difficult concept. Because we can assign a symbol to represent `infinity'
 and can manipulate that symbol according to specified rules, we assume
 corresponding ``infinite'' entities can exist in practice.  This is
 questionable\footnote{Our discussion here follows EKS, with the addition
 of supporting philosophical material and references.}. Furthermore, as we
have already indicated, such infinities lead to severe calculational problems
in the mathematical modelling of ensembles of universes or universe domains,
blocking any meaningful application of probability calculus.

It is very helpful to recognize at the outset that there are two different
concepts of ``the infinite'' which are often used: The {\it metaphysical}
infinite,
which designates wholeness, perfection, self-sufficiency; and the {\it
mathematical}
infinite, which represents
 that which is without limit (Moore 1990, pp.1-2, 34-
44; Bracken 1995, p.142, n.12). Here we are concerned with the
mathematical infinite\footnote{For a fascinating and very
readable, but somewhat eccentric, recent history of mathematical
infinity and its connections with key mathematical developments,
see David Foster Wallace (2003)}. But, now there are really two
basic categories of the mathematical infinite: The potential or
conceptual infinite and the actual, or realised, infinite. This
distinction goes back to rather diffuse but very relevant
discussions by Aristotle in his {\it Physics} and  his {\it
Metaphysics}. Basically, the potential or conceptual infinite
refers to a process or set conceptually defined so that it has no
limit to it -- it goes on and on, e.g. the integers. The concept
defining the set or process is  without a bound or limit, and
open, i. e. it does not repeat or retrace what is already produced
or counted. The actual, or realized, infinite would be a concrete
real object or entity, or set of objects, which is open and has no
limit to its specifications (in space, time, number of components,
etc.), no definite upper bound. Aristotle and many others since
have argued that, though there are many examples of potential or
conceptual infinities, actual realised infinities are not possible
as applied to entities or groups of entities.\footnote{Bracken
(1995, pp.11-24) gives a recent critical summary of Aristotle's
treatment of these issues, and their later use by Thomas Aquinas,
Schelling and Heidegger (Bracken, pp.25-51).}

There is no conceptual problem with an infinite set -- countable or
uncountable -- of \textit{possible} or \textit{conceivable } universes.
However, as David Hilbert (1964) points out,
the presumed existence of the actually infinite directly or indirectly
leads to well-recognised unresolvable contradictions in set theory (e. g.,
the Russell paradox, involving the set of all sets which do not contain
themselves, which by definition must both be a member of itself and not a
member of itself!), and thus in the definitions and deductive foundations
of mathematics itself (Hilbert, pp.141-142).

Hilbert's basic position is that ``Just as operations with the
infinitely small were replaced by operations with the finite which yielded
exactly the same results . . ., so in general must deductive methods based
on the infinite be replaced by finite procedures which yield exactly the
same results.\textquotedblright\ (p.135) He strongly maintains that ``the
infinite is nowhere to be found in reality, no matter what experiences,
observations, and knowledge are appealed to.\textquotedblright\ (p.142,
see also pp.136-137) Further on he remarks, ``Material logical deduction is
indispensable. It deceives us only when we form arbitrary abstract definitions,
especially those which involve infinitely many objects. In such cases we have
illegitimately used material logical deduction; i.e., we have not paid
sufficient attention to the preconditions necessary for its valid
use.\textquotedblright\
(p.142). Hilbert concludes, ``Our principal result is that the infinite is
 nowhere to be found in reality. It neither
exists in nature nor provides a legitimate basis for rational thought . . .
The role that remains for the infinite to play is solely that of an idea . .
. which transcends all experience and which completes the concrete as a
totality . . .'' (Hilbert, p.151).

\subsection{Arguments against Actual Infinity}

What are we to make of these intuitions and arguments? There are many
mathematicians and philosophers who espouse them. There are also
a large number who maintain that they are flawed. From the point of
view of cosmology itself, it would be very helpful if we could trust
in the conclusion that an actualized mathematical infinity is physically
impossible. For then this would provide a constraint on the scenarios
we use in cosmology, and assure us that probability calculations using them
could be successfully pursued. If, instead, there emerges a clear
indication that actual infinite sets are possible,
that would be mathematically disappointing. However, it still would be
an important conclusion, providing guidance and reassurance
in our quest to understand not only our observable universe, but the
universe or multiverse as whole, even though we will never have
\textit{direct} access to all of it (see Section 7, below).

It is not possible to explore this issue conclusively here. Philosophers
and philosophers of mathematics and science have proposed
many arguments against the possibility
of realized or actual mathematical infinities, and many
others, arguments in their favor. A careful critical review
is far beyond the scope of this paper. We need, however, to go
beyond the general and somewhat unfocused reasonings of Aristotle,
Hilbert and others we have summarized in the introduction to
this section. Thus, we shall briefly but more carefully present
several arguments against actual mathematical infinities which we
consider the strongest. Then in the next section we shall
explore in detail some of the mathematical and physical reasons
for which we should avoid admitting actual infinities.

We begin by proposing several key definitions. An \emph{actually
existing set} is one which has concrete physical status in our
extra-mental, intersubjective experience, and each of whose
members has a determinate, phenomenally supported status in our
experience, distinct from other members, with physically
characterisable features and integrity (e. g., a certain mass  or
energy, etc.). If the members of the set are not distinct or
determinate, then the set is either not well-defined or is not
actually existing.\footnote{In quantum
theory, the members of a set (e. g. particles) may be virtual,
going into and out of existence, but at any one time there are
only ``so many.'' Furthermore, there is a number operator, even
though the particles themselves cannot be physically
distinguished.} Any such actually existing set, and the members
constituting it, is contingent. They depend on something else for
being the way they are. It came into existence at a certain
moment, or within a certain series of moments, as the result of a
certain process, and eventually dissipates or dissolves, gradually
changing into something else over time. This is a pervasive
feature of our experience and of our scientific investigation of
physical reality.

An \emph{infinite set} is, as we have already said, one the number
of whose members is open, indeterminate and unbounded. By
``indeterminate'' we mean unspecified in terms of a definite
number. Infinity, strictly speaking is not a number in any usual
sense -- it is beyond all specific numbers which might be assigned
to a set or system -- it is simply the code-word for
\textquotedblleft it continues without end\textquotedblright. This
definition reflects how the term ``infinity'' is used in
mathematical physics, and in most of mathematics. The key point is
that there is no specific number which can be assigned to the
number of elements in an infinite set. One could say that the number of
its elements is one of Cantor's transfinite numbers, but those are
not numbers which specify a determinate bound.

An \emph{infinite physically existing set} is an infinite set
which is also an actually existing set -- in other words it
possesses an infinity of really existing objects. The number of
its objects is therefore unbounded and indeterminate. That means
in essence that, though it is unbounded and indeterminate in
number, it is nevertheless physically realized as complete. One
should note at this point that the definition is contradictory,
not from a logical or mathematical point of view (see Stoeger
2004), but from the point of view of physics and metaphysics. Can
what is essentially indeterminate and unbounded be physically or
really complete, which seems to imply ``bounded''? Can anything
which physically exists be completely unbounded? It is clear from
our definition of infinity that it is not a specific number we can
determine. Not only is it beyond any number we can specify or
conceive. It is unboundedly large and indeterminate. But an actual
infinity is conceived as extra-mentally instantiated and therefore
as completed. That means it must be determinate and in some
definite sense bounded. But this contradicts the definition of
what infinity is. Something cannot be bounded and determinate, and
unbounded and indeterminately large, at the same time. Therefore,
an actual physically realized infinity is not possible.

This appears, on the face of it, to be a compelling conceptual
argument against an actual really existing infinite set
of objects, whether they be universes, or something else.
The only way to counter it would be to show that an
indeterminate unboundedly large set is physically
realisable. Conceptually this seems to be impossible, just
from the point of view of what we mean by physically
concrete or actual, which seems to demand specifications
and some boundedness. Something which is unboundedly large,
and therefore not specifiable or determinate in quantity or
extent, is not materially or physically realisable. It is
not just that we are incapable of knowing an actual
infinity. It seems to involve a definite physical
impossibility -- the unbounded indeterminateness essential
to infinity is inconsistent with what it means to be
physically instantiated.

Another way of putting this is that the definition of infinity is
an issue in mathematics, not in physics. The problem arises in
linking the well-defined mathematical concept of infinity with
attempts at its realisation in physics. Realisations in physics
must have some determinateness -- but infinity as such is not
determinate!

This is recognized implicitly in scientific and applied
mathematics practice. Whenever
infinite values of physical parameters arise in physics --
such as in the case of singularities -- we can be reasonably
sure, as is often indicated, that there has been a
breakdown in our models. An achieved infinity in any physical
parameter (temperature, density, spatial curvature) is almost
certainly \textit{not} a possible outcome of any physical
process -- simply because it is unboundly large, indefinite
and indeterminate.

A second  supporting argument against realized infinity can be
constructed as follows. Since a realised infinite set of objects is
actually existing as a physical set, it is contingent and
therefore must have come into existence by some generating
process.\footnote{When contemplating mathematical concepts, it is
debatable as to whether a procedure or process is needed. But we
are talking physics, and the issue is precisely whether or not the
concept is realisable in the physical sense.} Then there are two
possibilities: 1. it became an actual infinite set by some process
of successive addition; or 2. it was produced as an infinite set
all at once.

But 1. does not work, since one cannot achieve a physically
infinite set by successive addition -- we can never actually \textit{arrive}
at infinity that way (see Spitzer 2000 and Stoeger 2004, and references
therein). There
is no physical process or procedure we can in principle implement
to complete such a set -- they are simply incompletable.
Some will concede that we can never physically
arrive at infinity in a finite time (see Smith 1993), but maintain
we can do so in an infinite time. But then we have the same problem
again with time -- that, for that to happen, we must \textit{complete}
an infinite number of events. But that seems to contradict the essentially
unbounded and indeterminate character of ``infinite time.''

So that leaves us with possibility 2., that the infinite set was
physically produced all at once. This is the one possibility
Bertrand Russell admits (Russell 1960). But, to produce an
infinite realized set of physical objects all at once requires a
process which makes an actually infinite amount of mass-energy
available. Again this must be a real complete, specified, infinite
amount of mass-energy. But this seems conceptually contradictory
again, for similar reasons.

Furthermore, the specification ``all at once'' demands
simultaneity, which is totally coordinate dependent. What is
simultaneous with respect to one coordinate system is not
simultaneous with respect to another. Thus, there is no assurance
to begin with that one can avoid the temporal completion problem
with 1. above; indeed one cannot do so with respect to all coordinate
systems. Therefore, once again here on several counts it seems
that a really infinite set of physical objects is not realisable
or actualizable.

These arguments underscore the fact that the problem with a
realised infinity is not primarily physical in the usual sense --
it is primarily a conceptual or philosophical problem.
``Infinity'' as it is mathematically conceived and used, is not
the sort of property that can be physically realised in an entity,
an object, or a system, like a definite number can. It is
indeterminately large, and really refers to a process rather than
to an entity (Bracken 1995, pp. 11-24). And the process it refers
to has no term or completion specified. No physically meaningful
parameter really possesses an infinite value. It is true that
cosmologists and physicists use infinities in ways which seem to
border on realised infinities, such as an infinite number of
points in a line segment, an infinite number of directions from
any point in three-space, or an infinite dimensional Hilbert
space.\footnote{We thank an anonymous referee for pointing out
these examples.} However, these are potential infinities,
indicating possible directions, locations or states that could be
taken or occupied. In no case are they all realised, occupied or
taken by distinguishable, really existing entities.

Finally, it is worth emphasizing that actual physically realized
infinities lead to a variety of apparently irresolvably
paradoxical, if not contradictory results (see Craig 1993) in
thought experiments, such as those involving adding to and
borrowing books from a really infinite library, or putting up new
guests in an already fully occupied hotel of an infinite number of
rooms. In fact, just the notion of a completed infinite set seems
to underlie some of the disturbing paradoxes of set theory (see
Craig 1993 for a brief discussion and references).

This issue is distinct from the difference between an ontologically realised
infinity or an epistemologically realised infinity. What we have
presented above seems to undermine the possibility of the former,
at least as a physical possibility. But it is a separate question
whether or not, granted the existence of a physically realized infinity,
it could ever be known or specified as such in a completed and determinate
way.

\subsection{Actual Infinities in Cosmology?}

Whether or not actual infinities are possible, they certainly need
to be avoided on the physical level, in order to make progress in
studying multiverses. As we have already discussed, actual
infinities lead to irresolvable problems in making probability
calculations; and their existence or non-existence is certainly
not observationally provable. They are an untestable proposal.

In the physical universe spatial infinities can be avoided by
compact spatial sections, either resultant from positive spatial
curvature or from choice of compact topologies in universes that
have zero or negative spatial curvature, (for example FLRW\ flat
and open universes can have finite rather than infinite spatial
sections). We argue that the theoretically possible infinite space
sections of many cosmologies at a given time are simply
unattainable in practice - they are a theoretical idea that cannot
be realised. It is certainly unprovable that they exist, if they
do. However one can potentially get evidence against such
infinities - if either it is observationally proven that we live
in a a `small universe', where we have already seen round the
universe because the spatial sections are compact on a scale
smaller than the Hubble scale (Lachieze-Ray and Luminet
1995);\footnote{There are some observational indications that this
could be so (see Sec.(\ref{small})), but they are far from
definitive.} or if we prove that the spatial curvature of the
best-fit FLRW universe model is positive, which necessarily
implies closed spatial sections (see Sec.\ref{small} below).

Future infinite time also is never realised: rather the situation
is that whatever time we reach, there is always more time
available.\footnote{Obviously this does not mean that we reject
standard Big Bang cosmology -- rejecting the really spatially
infinite universes as unrealizable does not undermine the
observational adequacy of these models, nor the essence of the Big
Bang scenario, even in the cases of those which are flat or open.
It just indicates that these models are incomplete, which we
already recognized.}  Much the same applies to claims of a past
infinity of time:\ there may be unbounded time available in the
past in principle, but in what sense can it be attained in
practice? The arguments against an infinite past time are strong -
it is simply not constructible in terms of events or instants of
time, besides being conceptually indefinite.\footnote{One way out
would be, as quite a bit of work in quantum cosmology seems to
indicate, to have time originating or emerging from the
quantum-gravity dominated primordial substrate only ``later.'' In
other words, there would have been a ``time'' or an epoch before
time as such emerged. Past time would then be finite, as seems to
be demanded by philosophical arguments, and yet the timeless
primordial state could have lasted ``forever,'' whatever that
would mean. This possibility avoids the problem of
constructibility.}

The same problem of
a realised infinity appears in considering supposed ensembles of really existing
universes. Aside from the strictly philosophical issues we have discussed
above, conceiving of an ensemble of many `really
existing' universes that are totally causally disjoint from our own, and how
that could come into being presents a severe challenge to cosmologists. There
are two fundamental reasons for this.
First, specifying the geometry of a generic universe requires an infinite
amount of information because the quantities in $\mathcal{P}_{\mathrm{\
geometry}}$ are fields on spacetime, in general requiring specification at
each point (or equivalently, an infinite number of Fourier coefficients) -
they will almost always not be algorithmically compressible. This greatly
aggravates all the problems regarding infinity and the ensemble itself. Only in
highly symmetric cases, like the FLRW solutions, does this data reduce to a
finite number of parameters. One can suggest that a statistical description
would suffice, where a finite set of numbers describes the statistics of the
solution, rather than giving a full description. Whether this suffices to
describe adequately an ensemble where `all that can happen, happens' is a
moot point. We suggest not, for the simple reason that there is no guarantee
that all possible models will be included in any known statistical description.
That
assumption is a major restriction on what is assumed to be possible.

Secondly, if many universes in the ensemble themselves really have infinite
spatial extent and contain an infinite amount of matter, that entails certain
 deeply paradoxical conclusions (Ellis and Brundrit 1979). To conceive
of the physical creation of an infinite set of universes (most requiring an
infinite amount of information for their prescription, and many of which
will themselves be spatially infinite) is at least an order of magnitude
more difficult than specifying an existent infinitude of finitely
specifiable objects.

The phrase `everything that can exist, exists' implies such an infinitude,
but glosses over all the profound difficulties implied. One should note here
particularly that problems arise in this context in terms of the continuum
assigned by classical theories to physical quantities and indeed to
spacetime itself. Suppose for example that we identify corresponding times
in the models in an ensemble and then assume that \textit{all} values of the
density parameter occur at each spatial point at that time. On the one hand, because of the
real number continuum, this is an uncountably infinite set of models -- and,
as we have already seen,
genuine existence of such an uncountable infinitude is highly problematic.
But on the other hand, if the set of realised models is either finite or
countably infinite, then almost all possible models are not realised -- the
ensemble represents a set of measure zero in the set of possible universes.
Either way the situation is distinctly uncomfortable.

 However, we might try
to argue around this by a discretization argument:\ maybe
differences in some parameter of less than say $10^{-10}$ are
unobservable, so we can replace the continuum version by a
discretised one, and perhaps some such discretisation is forced on
us by quantum theory - indeed this is a conclusion that follows
from loop quantum gravity, and is assumed by many to be the case
whether loop quantum gravity is the best theory of quantum gravity
or not. That solves the `ultraviolet divergence' associated with
the small-scale continuum, but not the `infrared divergence'
associated with supposed infinite distances, infinite times, and
infinite values of parameters describing cosmologies.

Even within the restricted set of FLRW\ models, the problem of
realised infinities is profoundly troubling: if all that is
possible in this restricted subset happens, we have multiple
infinities of realised universes in the ensemble. First, there are
an infinite number of possible spatial topologies in the negative
curvature case (see e.g. Lachieze-Ray and Luminet 1995), so an
infinite number of ways that universes which are locally
equivalent can differ globally. Second, even though the geometry
is so simple, the uncountable continuum of numbers plays a
devastating role locally: is it really conceivable that FLRW
universes actually occur with \textit{all} values independently of
both the cosmological constant and the gravitational constant, and
also all values of the Hubble constant, at the instant when the
density parameter takes the value 0.97? This gives 3 separate
uncountably infinite aspects of the ensemble of universes that are
supposed to exist. Again, the problem would be allayed if
spacetime is quantized at the Planck level, as suggested for
example by loop quantum gravity. In that case one can argue that
all physical quantities also are quantized, and the uncountable
infinities of the real line get transmuted into finite numbers in
any finite interval -- a much better situation. We believe that
this is a physically reasonable assumption to make, thus softening
a major problem for many ensemble proposals. But the intervals are
still infinite for many parameters in the possibility space.
Reducing the uncountably infinite to countably infinite does not
in the end resolve the problem of infinities in these ensembles.
It is still an extraordinarily extravagant proposal, and, as we
have just discussed, seems to founder in the face of careful
conceptual analysis.

The argument given so far is based in the nature of the
application of mathematics to the description of physical reality.
We believe that it carries considerable weight, even though the
ultimate nature of the mathematics-physics connection is one of
the great philosophical puzzles. It is important to recognize,
however, that arguments regarding problems with realised infinity
arise from the physics side, independently of the mathematical and
conceptual consideration we have so far emphasized.

On one hand, broad quantum theoretical considerations suggest that space-time
may be discrete at the Planck scale, and some specific quantum gravity
models indeed have been shown to incorporate this feature when examined
in detail. If this is so, not only does it remove the real number line
as a {\it physics} construct, but it {\it inter alia} has the potential
to remove the ultraviolet divergences that otherwise plague field theory --
a major bonus.

On the other hand, it has been known for a long time that there are
significant problems with putting boundary conditions for physical
theories actually at infinity. It was for this reason that Einstein
preferred to consider universe models with compact spatial sections
(thus removing the occurrence of spatial infinity in these models).
This was a major motivation for his static universe model proposed
in 1917, which necessarily has compact space sections. John Wheeler
picked up this theme, and wrote about it extensively in his book
{\it Einstein's Vision} (1968). Subsequently, the book {\it Gravitation}
by Misner, Thorne and Wheeler (1973) only considered spatially
compact, positively curved universe models in the main text. Those
with flat and negative spatial curvatures where relegated to a
subsection on ``Other Models.''

Thus, this concern regarding infinity has a substantial physics
provenance, independent of Hilbert's mathematical arguments and
philosophical considerations. It recurs in present day
speculations on higher dimensional theories, where the higher
dimensions are in many cases assumed to be compact, as in the
original Kaluza-Klein theories. Various researchers have then
commented that ``dimensional democracy'' suggests all spatial
sections should be compact, unless one has some good physical
reason why those dimensions that remained small are compact while
those that have expanded to large sizes are not. Hence we believe
there is substantial support from physics itself for the idea that
the universe may have compact spatial sections, thus also avoiding
infra-red divergences -- even though this may result in
``non-standard'' topologies for its spatial sections. Such
topologies are commonplace in string theory and in M-theory --
indeed they are essential to their nature.

\section{On the origin of ensembles}

Ensembles have been envisaged both as resulting from a single causal
process, and as simply consisting of discrete entities. We discuss these two
cases in turn, and then show that they are ultimately not distinguishable
from each other.

\subsection{Processes Naturally Producing Ensembles}

Over the past 15 or 20 years, many researchers investigating the very early
universe have proposed processes at or near the Planck era which would
generate a really existing ensemble of expanding universe domains, one of
which is our own observable universe. In fact, their work has provided both
the context and stimulus for our discussions in this paper. Each of these
processes essentially selects a really existing ensemble
from a set of possible universes, often leading to a
proposal for a natural definition of a probability distribution on the
space of possible universes. Here we briefly describe some of these,
and comment on how they fit within the framework we have been
discussing.

The earliest explicit proposal for an ensemble of universes or universe
domains was by Vilenkin (1983). Andrei Linde's chaotic inflationary proposal
(Linde 1983, 1990, 2003)  is
one of the best known scenarios of this type. The scalar field (inflaton) in
these scenarios drives inflation and leads to the generation of a large
number of causally disconnected regions of the Universe. This process is
capable of generating a really existing ensemble of expanding FLRW-like
regions, one of which may be our own observable universe region, situated in
a much larger universe that is inhomogeneous on the largest scales. No FLRW\
approximation is possible globally; rather there are many FLRW-like
sub-domains of a single fractal universe. These domains can be very
different from one another, and can be modelled locally by FLRW cosmologies
with different parameters.

Vilenkin, Linde and others have applied a stochastic approach to inflation
(Vilenkin 1983, Starobinsky 1986, Linde, \textit{et al.} 1994, Vilenkin 1995, Garriga and
Vilenkin 2001, Linde 2003), through which probability distributions can be
derived from inflaton potentials along with the usual cosmological equations
(the Friedmann equation and the Klein-Gordon equation for the inflaton) and
the slow-roll approximation for the inflationary era. A detailed example of
this approach, in which specific probability distributions are derived from
a Langevin-type equation describing the stochastic behaviour of the inflaton
over horizon-sized regions before inflation begins, is given in Linde and
Mezhlumian (2003) and in Linde \textit{et al.} (1994). The probability
distributions determined in this way generally are functions of the inflaton
potential.

As we mentioned in the introduction, over the past few years
considerable progress has been achieved by theorists in developing
flux stabilized, compactified, non-supersymmetric solutions to
superstring/M theory which possess an enormous number of vacua
(Susskind 2003, Kachru, et al. 2003, and references therein). Each
of these vacua has the potential for becoming a separate universe
or universe domain, with a non-zero cosmological constant. As such
it is relatively easy to initiate inflation in many of them.
Furthermore, the dynamics leading to these vacua also generate
different values of the some of the other cosmological and
physical parameters, and enable a statistical treatment of string
vacua themselves.

These kinds of scenario suggests how overarching physics, or a
\textquotedblleft law of laws\textquotedblright (represented by the inflaton
field and its potential), can lead to a really existing ensemble of many
very different FLRW-like regions of a larger Universe. However these
proposals rely on extrapolations of presently known physics to realms far
beyond where its reliability is assured. They also employ inflaton
potentials which as yet have no connection to the particle physics we know
at lower energies. And these proposals are not directly observationally
testable -- we have no astronomical evidence that the supposed other FLRW-like
regions exist, and indeed do not expect to ever attain such evidence. Thus they
remain theoretically based proposals rather than
provisionally acceptable models -- much less established fact. There remains
additionally the difficult problem of
infinities, which we have just discussed: eternal inflation with its
 continual reproduction of different
inflating domains of the Universe is claimed to lead to an
infinite number of universes of each particular type (Linde,
private communication). How can one deal with these infinities in
terms of distribution functions and an adequate measure? As we
have pointed out above, there is a philosophical problem
surrounding a realised infinite set of any kind. In this case the
infinities of really existent FLRW-like domains derive from the
assumed initial infinite flat (or open) space sections - and we
have already pointed out the problems in assuming such space
sections are actually realised. If this is correct, then at the
very least these proposals must be modified so that they generate
a finite number of universes or universe domains.

Finally, from the point of view of the ensemble of all possible universes
often invoked in discussions of multiverses, all possible inflaton
potentials should be considered, as well as all solutions to all those
potentials. They should all be represented in $\mathcal{M}$, which will include
chaotic inflationary models which are stationary as well as those which are
non-stationary. Many of these potentials may yield ensembles which are
uninteresting as far as the emergence of life is concerned, but some will be
bio-friendly.

In EKS we have briefly reviewed various proposals for probability
distributions of the cosmological constant over ensembles of
universe domains generated by the same inflaton potential,
particularly those of Weinberg (2000) and Garriga and Vilenkin
(2000, 2001). We shall not revisit this work here, except to
mention the strong anthropic constraints on values of the
cosmological constant, which is the primary reason for interest in
this case. Galaxy formation is only possible for a narrow range of
values of the cosmological constant, $\Lambda$, around $\Lambda =
0$ (one order of magnitude - hugely smaller than the 120 orders of
magnitude predicted by quantum field theory as its natural value).

\subsection{Testability of these proposals}
 In his popular book {\it Our
Cosmic Habitat} Martin Rees (Rees 2001b, pp. 175ff) uses this
narrow range of bio-friendly values of $\Lambda$ to propose a
preliminary test which he claims could rule out the multiverse
explanation of fine-tuning for certain parameters like $\Lambda$.
This is what might be called a ``speciality argument.'' According
to Rees, if ``our universe turns out to be {\it even more
specially} tuned than our presence requires,'' the existence of a
multiverse to explain such ``over- tuning'' would be refuted. The
argument itself goes this way. Naive quantum physics expects
$\Lambda$ to be very large. But our presence in the universe
requires it to be very small, small enough so that galaxies and
stars can form. Thus, in our universe $\Lambda$ must obviously be
below that galaxy-forming threshold. This explains the observed
very low value of $\Lambda$ as a selection effect in an existing
ensemble of universes. Although the probability of selecting at
random a universe with a small $\Lambda$ is very small, it becomes
large when we add the prior that life exists. Now, in any universe
in which life exists, we would not expect $\Lambda$ to be too far
below this threshold. Otherwise it would be more fine-tuned than
needed. In fact, data presently indicates that $\Lambda$ is not
too far below the threshold, and thus our universe is not markedly
more special than it needs to be, as far as $\Lambda$ is
concerned. Consequently, explaining its fine-tuning by assuming a
really existing multiverse is acceptable. Rees suggests that the
same argument can be applied to other parameters.

Is this argument compelling? As Hartle (2004) has pointed out, for
the first stage to be useful, we need an {\it a priori}
distribution for values of $\Lambda$ that is very broad, combined
with a very narrow set of values that allow for life. These values
should be centred far from the most probable {\it a priori}
values. This is indeed the case if we suppose a very broad
Gaussian distribution for $\Lambda$ centred at a very large value,
as suggested by quantum field theory. Then, regarding the second
stage of the argument, the values allowing for life fall within a
very narrow band centred at zero, as implied by astrophysics.
Because the biophilic range is narrow, the {\it a priori}
probability for $\Lambda$ will not vary significantly in this
range. Thus, it is equally likely to take any value. Thus a {\it
uniform probability assumption} will be reasonably well satisfied
within the biophilic range of $\Lambda$.

As regards this second stage of the argument, because of the uniform probability
assumption it is not clear why the expected values for the existence of
galaxies should pile up near the biological limit. Indeed, one might expect the
probability of the existence of galaxies to be maximal at the centre of the
biophilic range rather than at the edges (this probability drops to zero at the
edges, because it vanishes outside -- hence the likelihood of existence of
galaxies at the threshold itself should be very small). Thus there is no
justification on this basis for ruling out a multiverse with any specific value for
$\Lambda$ within that range.
As long as the range of values of a parameter like $\Lambda$ is
not a zero-measure set of the ensemble, there is a non-zero
probability of choosing a universe within it. In that case, there
is no solid justification for ruling out a multiverse and so no
real testability of the multiverse proposal. All we can really say
is that we would be less likely to find ourselves in a universe
with a $\Lambda$ in that range in that particular ensemble. Indeed
no probability argument can conclusively \emph{disprove} any
specific result - all it can state is that the result is
improbable - but that statement only makes sense if the result is
possible! What is actually meant by ``more specially tuned than
necessary for our existence''?  In the end, any particular choice
of a life-allowing universe will be more specially tuned for
something. In our view ``tuning'' refers to parameters selected
such that the model falls into a certain class, e.g.,
life-allowing. Any additional tuning would then just be the
selection of sub-classes, and, after all, any particular model is
``over-tuned'' in such a way as to select uniquely the sub-class
which contains only itself. Rees's argument seems to imply that
$\Lambda$ close to zero would be an over-tuned case, while
$\Lambda$ close to the cut-off value would not be. However, would
the reversed viewpoint be not just as legitimate?

Rees's argument strongly builds on the predictions of quantum
physics --- a probability distribution peaked at very high values
for $\Lambda$. Taking into account the unknown relation between
general relativity and quantum physics we should treat the problem
as a multiple hypothesis testing problem: The multiverse scenario
can be true or false, and so can the quantum prediction for high
values of $\Lambda$. An observed low value of $\Lambda$ would then
strongly question the predictions for $\Lambda$, but say nothing
about the multiverse scenario. We conclude that any observed value
of $\Lambda$ does not rule out the multiverse scenario. It also
seems questionable whether the life-allowing values for $\Lambda$
can be classified just by a simple cut-off value. It should be
expected that there are more subtle and yet unknown constraints.
Observing a cosmological constant far from the cut-off value might
then just be the result of some unknown constraints.

Finally, probability arguments simply don't apply if there is
indeed only one universe - their very use assumes a multiverse
exists. There might exist only one universe which just happens to
have the observed value of $\Lambda$; then probabilistic arguments
will simply not apply. Thus what we are being offered here is not
in fact a proof a multiverse exists, but rather a consistency
check as regards the nature of the proposed multiverse. It is a
proposal for a necessary but not sufficient condition for its
existence. As emphasized above, we do not even believe it is a
necessary condition; rather it is a plausibility indicator.

\subsection{The existence of regularities}

Consider now a genuine multiverse. Why should there be any regularity at all
in the properties of universes in such an ensemble, where the universes are
completely disconnected from each other? If there are such regularities and
specific resulting properties, this suggests a mechanism creating that
family of universes, and hence a causal link to a higher domain which is the
seat of processes leading to these regularities. This in turn means that the
individual universes making up the ensemble are not actually independent of
each other. They are, instead, products of a single process, or meta-process, as
in the case
of chaotic inflation. A common generating mechanism is clearly a causal
connection, even if not situated in a single connected spacetime -- and some
such mechanism is needed if all the universes in an ensemble have the same
class of properties, for example being governed by the same physical laws or
meta-laws.

The point then is that, as emphasized when we considered how one can
describe ensembles, any multiverse with regular properties that we can
characterise systematically is necessarily of this kind. If it did not have
regularities of properties across the class of universes included in the
ensemble, we could not even describe it, much less calculate any properties
or even characterise a distribution function.

Thus in the end the idea of a completely disconnected multiverse
with regular properties but without a common causal mechanism of
some kind is not viable. There must necessarily be some
pre-realisation causal mechanism at work determining the
properties of the universes in the ensemble. What are claimed to
be totally disjoint universes must in some sense indeed be
causally connected together, albeit in some pre-physics or
meta-physical domain that is causally effective in determining the
common properties of the universes in the multiverse. This is
directly related to the two key issues we highlighted above in
Sections 2 and 3, respectively, namely how does the possibility
space originate, and where does the distribution function that
characterises realised models come from?

From these considerations, we see that we definitely need to
explain (for Issue 2) what particular cosmogonic generating
process or meta-law pre-exists, and how that process or meta-law
was selected from those that are possible. Obviously an infinite
regress lurks in the wings. Though intermediate scientific answers
to these questions can in principle be given, it is clear that no
ultimate scientific foundation can be provided.

Furthermore, we honestly have to admit that any proposal for a
particular cosmic generating process or principle we establish as
underlying our actually existing ensemble of universe domains or
universes, after testing and validation (see Section 7 below),
will always be at best provisional and imperfect: we will never be
able to definitively determine its nature or properties. The
actually existing cosmic ensemble may in fact be much, much larger
-- or much, much smaller -- than the one our physics at any given
time describes, and embody quite different generating processes
and principles than the ones we provisionally settle upon. This is
particularly true as we shall never have direct access to the
ensemble we propose, or to the underlying process or potential
upon which its existence relies (see Section 7 below),
 nor indeed to the full range of physics that may be involved.

\subsection{The existence of possibilities}
Turning to the prior question (Issue 1, see Section 2.1), what
determines the space of all possible universes, from which a
really existing universe or an  ensemble of universes or universe
domains is drawn, we find ourselves in even much more uncertain
waters. This is particularly difficult when we demand some basic
meta-principle which delimits the set of possibilities. Where
would such a principle originate? The only two secure grounds for
determining possibility are existence ("ab esse ad posse valet
illatio") and freedom from internal contradiction. The first
really does not help us at all in exploring the boundaries of the
possible. The second leaves enormous unexplored, and probably
unexplorable, territory. There are almost certainly realms of the
possible which we cannot even imagine. But at the same time, there
may be, as we have already mentioned, universes we presently think
are possible which are not. We really do not have secure grounds
for determining the limits of possibility in this expanded cosmic
context. We simply do not have enough theoretical knowledge to
describe and delimit reliably the realm of the possible, and it is
very doubtful we shall ever have.

\section{Testability and Existence}

The issue of evidence and testing has already been briefly
mentioned. This is at the heart of whether an ensemble or multiverse
proposal should be regarded as physics or as metaphysics.

\subsection{Evidence and existence}

Given all the possibilities discussed here, which specific kind of ensemble
is claimed to exist? Given a specific such claim, how can one show that this
is the particular ensemble that exists rather than all the other
possibilities?

There is no direct evidence of existence of the claimed other universe
regions in an ensemble, nor can there be any, for they lie beyond the visual
horizon; most will even be beyond the particle horizon, so there is no
causal connection with them; and in the case of a true multiverse, there is
not even the possibility of any indirect causal connection - the
universes are then completely disjoint and nothing that happens in any one
of them is causally linked to what happens in any other one (see Section 6.2).
This lack of any causal connection in such multiverses really places them beyond
any scientific support -- there can be no direct or indirect evidence for the
existence of such systems. We may, of course, postulate the existence of such
a multiverse as a metaphysical assumption, but it would be a
metaphysical assumption without any further justifiability -- it would be
untestable and unsupportable by any direct or indeed indirect evidence.

And so, we concentrate on possible really existing multiverses in which there
is some common causal generating principle or process.
What weight does a claim of such existence carry in this case, when no
direct observational evidence can ever be available? The point is that there
is not just an issue of showing a multiverse exists. If this is a
scientific proposition one needs to be able to show which specific
multiverse exists; but there is no observational way to do this. Indeed if
you can't show \textit{which particular} one exists, it is doubtful you have
shown \textit{any} one exists. What does a claim for such existence mean in
this context? Gardner puts it this way: "There is not the slightest shred of
reliable evidence that there is any universe other than the one we are in. No
multiverse theory has so far provided a prediction that can be tested. As far
as we can tell, universes are not even as plentiful as even {\it two}
blackberries" (Gardner 2003). This contrasts strongly, for example, with
Deutsch's and Lewis's defence of the concept (Deutsch 1998, Lewis 2000).

\subsection{Fruitful Hypotheses and evidence}

There are, however, ways of justifying the existence of an entity, or entities,
like a
multiverse, even though we have no direct observations of it. Arguably the
most compelling framework within which to discuss testability is that of
\textquotedblleft retroduction\textquotedblright\ or \textquotedblleft
abduction\textquotedblright\ which was first described in detail by
C.S.Peirce. Ernan McMullin (1992) has convincingly demonstrated that
retroduction is the rational process by which scientific conclusions are
most fruitfully reached. On the basis of what researchers know, they
construct imaginative hypotheses, which are then used to probe and to
describe the phenomena in deeper and more adequate
ways than before. As they do so,
they will modify or even replace the original hypotheses, in order to make
them more fruitful and more precise in what they reveal and explain. The
hypotheses themselves may often presume the existence of certain hidden
properties or entities (like multiverses!) which are fundamental to the
explanatory power they possess. As these hypotheses become
more and more fruitful in revealing and explaining the natural phenomena
they investigate, and their inter-relationships, and more central to
scientific research in a given discipline, they become more
and more reliable accounts of the reality they purport to model or describe.
Even if some of the hidden properties or entities they postulate are never
directly detected or observed, the success of the hypotheses indirectly
leads us to affirm that something like them must exist.\footnote{In light of
discussions by McMullin elsewhere (McMullin 1993, pp. 381-382) more care and
precision is needed here. He recommends separating explanation from
proof of existence: ``In science, the adequacy of a theoretical explanation is
often regarded as an adequate testimony to the existence of entities postulated
by the theory. But the debates that swirl around this issue among philosophers
(the issue of scientific realism, as philosophers call it) ought to warn us of
the risks of moving too easily from explanatory adequacy to truth-claims for the
theory itself. This sort of inference depends sensitively on the quality of the
explanation given, on the viability of alternatives, on our prior knowledge of
beings in the postulated category, and on other more complex factors.''}
 A cosmological example is the
inflaton supposed to underlie inflation.

Thus, from this point of view, the existence of an ensemble of
universes or universe domains would be a validly deduced -- if
still provisional -- scientific conclusion if this becomes a key
component of hypotheses which are successful and fruitful in the
long term. By an hypothesis which manifests long-term success and
fruitfulness we mean one that better enables us to make testable
predictions which are fulfilled, and provides a more thorough and
coherent explanation of phenomena we observe than competing
theories.\footnote{It is interesting to note that Rees (2001b, p.
172) hints at the use of a retroductive approach in cosmology, but
does not develop the idea as an argument in any detail.} Ernan
McMullin (1992; see also P. Allen 2001, p. 113) frames such
fruitfulness and success as:

a.  accounting for all the relevant data (empirical adequacy);

b. providing long-term explanatory success and stimulating fruitful lines of
further inquiry (theory fertility);

c. establishing the compatibility of previously disparate domains of phenomena
(unifying power);

d. manifesting consistency and correlation with other established
theories (theoretical coherence).

The relevant example here would be a fruitful theory relying on a
specific type of multiverse, all members of which would never be
directly detectable except one. But, since its postulated
existence renders the existence and the characteristic features of
our own universe ever more intelligible and coherent over a period
of time, this can be claimed to be evidence for
the multiverse's existence. If such indirect support for the
existence of a given multiverse is inadequate in the light of
other competing accounts, then from a scientific point of view all
we can do is to treat it as a speculative scenario needing further
development and requiring further fruitful application. Without
that, espousing the existence of a given multiverse as the
explanation for our life-bearing universe must surely be called
metaphysics, because belief in its existence will forever be a
matter of faith rather than proof or scientific support.

We do, of course, want to avoid sliding to the bottom of Rees' (2001b, p. 169)
slippery slope. In arriving at his conclusion that the existence of other
universes
is a scientific question, Rees (2001b, pp. 165-169) begins by considering first
galaxies which are
beyond the limits of present-day telescopes, and then galaxies which are beyond
our
visual horizon now, but will eventually come within it in the future. In both
cases these galaxies
are real and observable {\it in principle}. Therefore, they remain legitimate
objects of
scientific investigation. However, then he goes on to consider galaxies which
are forever unobservable,
but which emerged from the same Big Bang as ours did. And he concludes that,
though unobservable, they
are real, and by implication should be included as objects considered by
science. Other universes,
he argues, fall in the same category -- they are real, and therefore they should
fall within
the boundaries of scientific competency. As articulated this is indeed ``a
slippery slope'' argument -- it can
be used to place anything that we claim to be ``real'' within the natural
sciences -- unless
we strengthen it at several points.

First, Rees shifts the criterion from ``observable
in principle'' to being ``real.'' This is really an error. No matter how real an
object, process,
or relationship may be, if it is not observable in principle, or if there is not
at least indirect support
for its existence from the long-term success of the hypotheses in which it
figures, then it simply
falls outside serious scientific consideration. It may still temporarily play a
role in scientific
speculation, but, unless it receives some evidential support, that will not
last. In mentioning
that the forever unobservable galaxies
he is considering are produced by the same Big Bang as ours, Rees may be
intending to indicate that,
though unobservable, they share a common causal origin and therefore {\it
figure} in successful
hypotheses, as would be required by McMullin's retroductive inference discussed
above. But Rees
does not make that clear. Moving to other universes, the same
requirement holds. Thus, the slippery slope is avoided precisely by implementing
the "indirect
evidence by fruitful hypotheses" approach that a careful application of
retroduction requires.

Second, there is discontinuity in the argument as one moves from weaker and
weaker causal relation to none at all. The slippery slope becomes a vertical
precipice on one side of an unbridgeable gulf. An argument that relies on
incremental continuity does not apply in this case.

Thus, if we are continually evaluating our theories and
speculations with regard to their potential and actual fruitfulness in
revealing and explaining the world around us, then we shall avoid the lower
reaches of the slippery slope. The problem is that, in this case, the
multiverse hypothesis is very preliminary and will probably always remain
provisional.
This should not prevent us from
entertaining imaginative scenarios, but the retroductive process will subject
these
speculations to rigorous critique over time. The key issue then is to what
degree will the multiverse hypothesis become fruitful. Unfortunately, as it
stands
now, it is not, because it can be used to explain anything at all -- and
hence does not explain anything in particular. You cannot predict something
new from the hypothesis, but you can explain anything you already know. In
order for it to achieve some measure of scientific fruitfulness, there must
be an accumulation of at least indirect scientifically acceptable support
for one particular well-defined multiverse.
Indeed, from a purely evidential viewpoint, a multiverse with say $10^{120}$
identical copies of the one universe in which we actually live would be much
preferred over one with a vast variety of different universes, – for then the
probability of finding a universe like our own would be much higher. Such
ensembles are usually excluded because of some hidden assumptions about the
nature of the generating mechanism that creates the ensemble. But maybe that
mechanism is of a different kind than usually assumed - perhaps once it has
found a successful model universe, it then churns out innumerable identical
copies of the same universe.

In the end belief in a multiverse may always be just that -- a matter of faith,
namely faith that the logical arguments discussed here give the correct
answer in a situation where direct observational proof is unattainable and
the supposed underlying physics is untestable, unless we are able to point
to compelling reasons based on scientifically supportable evidence for a
particular specifiable multiverse or one of a narrowly defined class of
multiverses. One way in which this could be accomplished, as we have already
indicated, would be to find
accumulating direct or indirect evidence that a very definite inflaton
potential capable of generating a certain type of ensemble of universe
domains was operating in the very early universe, leading to the particular
physics that we observe now. Otherwise, there will be no way of ever knowing
which particular multiverse is realised, if any one is. We will always be
able to claim whatever we wish about such an ensemble, provided it includes
at least one universe that admits life.

\subsection{Observations and Physics}
One way one might make a reasonable claim for existence of a multiverse
would be if one could show its existence was a more or less inevitable
consequence of well-established physical laws and processes. Indeed, this is
essentially the claim that is made in the case of chaotic inflation. However
the problem is that the proposed underlying physics has not been tested, and
indeed may be untestable. There is no evidence that the postulated physics
is true in this universe, much less in some pre-existing metaspace that
might generate a multiverse.

Thus there are two further
requirements which must still be met, once we have proposed a viable
ensemble or multiverse theory. The first is to provide some credible link
between these vast extrapolations from presently known physics to physics in
which we have some confidence. The second is to provide some at least
indirect evidence that the scalar potentials, or other overarching cosmic
principles involved, really have been functioning in the very early
universe, or before its emergence. We do not at present fulfil either
requirement.

The issue is not just that the inflaton is not
identified and its potential untested by any observational means - it is
also that, for example, we are assuming quantum field theory remains valid
far beyond the domain where it has been tested, and where we have faith in that
extreme extrapolation despite all the unsolved problems at the foundation of
quantum theory, the divergences of quantum field theory, and the failure of
that theory to provide a satisfactory resolution of the cosmological
constant problem.

\subsection{Observations and disproof}\label{small}

Despite the gloomy prognosis given above, there are some specific
cases where the existence of a chaotic inflation (multi-domain)
type scenario can be disproved. These are when we either live in a
universe with compact spatial sections because they have positive
curvature, or in `small universe' where we have already seen right
round the universe (Ellis and Schreiber 1986, Lachieze-Ray and
Luminet 1995), for then the universe closes up on itself in a
single FLRW-like domain, and so no further such domains that are
causally connected to us in a single connected spacetime can
exist.

As regards the first case, the best combined astronomical data at
present (from the WMAP satellite together with number counts and
supernova observations) suggest that this is indeed the case: they
indicate that $\Omega_0 = 1.02 \pm 0.02$ at a 2-$\sigma$ level, on
the face of it favoring closed spatial sections and a spatially
finite universe. This data does not definitively rule out open
models, but it certainly should be taken seriously in an era of
`precision cosmology.'

As regards the `small universe' situation, this is in principle
observationally testable, and indeed it has been suggested that
the CBR\ power spectrum might already be giving us evidence that
this is indeed so, because of its lack of power on the largest
angular scales (Luminet et al, 2003). This proposal can be tested
in the future by searching for identical circles in the CMB sky
(Roukema, et al., 2004) and alignment of the CMB quadrupole and
octopole planes (Katz and Weeks 2004). Success in this endeavour
would disprove the usual chaotic inflationary scenario, but not a
true multiverse proposal, for that cannot be shown to be false by
any observation. Neither can it be shown to be true.

\section{Special or Generic?}

When we reflect on the recent history of cosmology, we become aware that
philosophical predilections have oscillated from assuming that the present
state of our universe is very special (made cosmologically precise in
contemporary cosmology as FLRW, or almost-FLRW, through the assumption of a
Cosmological
Principle -- see Bondi 1960 and Weinberg 1972, for example), requiring
very finely tuned initial conditions, to assuming it is generic, in the
sense that it has attained its present apparently special qualities through
the operation of standard physical processes on any of broad range of possible
initial conditions (e. g.,  the ``chaotic cosmology'' approach
of Misner (1968) and the now standard but incomplete inflationary
scenario pioneered by Guth (1980)). This
oscillation, or tension, has been described and discussed in detail, both in its
historical and in its contemporary manifestations, by McMullin
(1993) as a conflict or tension between two general types of principle --
anthropic-like principles, which recognize the special character of the
universe and tentatively presume that its origin must be in finely tuned
or specially chosen
initial conditions, and  ``cosmogonic indifference principles,'' or just
``indifference principles,'' which concentrate their search upon very
generic initial conditions upon which the laws of physics act to produce
the special cosmic configuration we now enjoy. As McMullin
portrays these two philosophical commitments, the anthropic-type preference
inevitably attempts to involve mind and teleology as essential to the
shaping of what emerges, whereas the indifference-type
preference studiously seeks to avoid any direct appeal to such
influences, relying instead completely upon the dynamisms (laws of
nature) inherent in and emerging from mass-energy itself.

McMullin (1993), in a compelling historical sketch, traces the preference
for the special and the teleologically suggestive from some of the earlier
strong anthropic principle formulations back through early Big Bang cosmology
to Clarke, Bentley, William Derham and John Ray in the 18th and 19th
centuries and Robert Boyle in the 17th century and ultimately back to
Plato and the Biblical stories of creation. The competing preference for
indifferent initial conditions and the operation of purely physical or
biological laws can be similarly followed back from the present appeal to
multiverses to slightly earlier inflationary scenarios and Misner's
chaotic cosmology program to steady state cosmological models and then
back through Darwin to Descartes and much earlier to the Greek atomists, such
as Empedocles, Diogenes Laertius and Leucippus. Neither of these historical
sequences involves clearly dependent philosophical influences, but
the underlying basic assumptions and preferences of each of the two sets of
thinkers and models are very similar, as are their controversies and
interactions with the representatives of the competing approach.

Certainly it has become clear that the present preference among theoretical
cosmologists for multiverse scenarios is the latest and most concerted
attempt to implement the indifference principle in the face of the
mounting evidence that, taken alone, our universe does require very
finely tuned initial conditions. The introduction of inflation was
similarly motivated, but has encountered some scepticism in this regard
with the growing sense that initiating inflation itself probably requires
special conditions (Penrose 1989; Ellis, et al. 2002).
The appeal to multiverses, though first seriously suggested fairly
early in this saga (by Dicke in 1961 and by Carter in 1968), has been
reasserted as this failure of other indifference principle implementations
seem more and more imminent. However, as we have just seen in our detailed
discussion of realised ensembles of universes or universe domains, they are
by no means unique, and accounting for
their existence requires an adequate generating process or principle, which
must explain the distribution function characterizing the ensemble.

Even though we are far from being able to connect specific types of ensembles
with
particular provisionally adequate cosmogonic generating processes in a
compelling
way, it is very possible that some fine-tuning of these processes may be
required to
mesh with the physical constraints we observe in our universe and at the same
time to produce a realised ensemble which enbraces it. This would initiate
another
oscillation between the two types of principles. Whether or not that occurs, it
is clear that the existence of a multiverse in itself does not support either
the indifference principle nor the anthropic-type principle. What would do so
would be the distribution function specifying the multiverse, and particularly
the physical, pre-physical or metaphysical process which generates
the multiverse with that distribution function, or range of distribution
functions.
Only an understanding of that process would ultimately determine
which principle is really basic.

Whatever the eventual outcome of future
investigations probing this problem, it is both curious and striking, as
McMullin
(1993, p. 385) comments, that ``the same challenge arises over and over.'' Fine-
tuning at one level is tentatively explained by some process at a more
fundamental
level which seems at first sight indifferent to any initial conditions. But then
further investigation reveals that that process really requires special
conditions,
which demands some fine-tuning. Meanwhile, ``the universe'' required for
understanding
and explanation ``keeps getting larger and larger.''

It might seem that these competing philosophical or metaphysical preferences --
for
what is either basically special or basically generic -- are choices without
scientific
or philosophical support. But that is an illusion. From what we have seen
already, there
is considerable physical and philosophical support for each preference -- some
of it
observational and some of it theoretical -- but there is no {\it adequate} or
{\it definitive} support for one over against the other. Thus, either preference
may
be supported in various ways philosophically and scientifically, but neither the
one
nor the other is {\it THE} scientific approach. For example, the emergent
universe
model of Ellis, Maartens, et al. (2003) has fine-tuned initial conditions, but
it
still could be a good model -- it may actually represent how the very early
history of our universe unfolded, even though it does not explain how the
special
initial conditions were set. [In fact it is not as fine-tuned as inflation with
$k = 0$, which requires ``infinite'' fine-tuning, while being ``asymptotic'' to
an Einstein static universe does not.]

The issue is not
 so much
which of the two principles or perspectives are correct -- both
seem to be important at different levels and in different heuristic and
explanatory
contexts. As far as we know, there has not been any resolution to the question
of
the epistemological or ontological status of either one. They function rather as
contrary heuristic preferences which have both intuitive and experiential
support.
Perhaps the real question is: Which is more fundamental? It is possible that,
from
the point of view of physics, the indifference principle is more fundamental --
relative to the explanations which are possible within the sciences -- whereas
from the
point of view of metaphysics, an anthropic-type fine-tuning principle is more
fundamental.

What does seem clear, in this regard, is that the effort to keep
explanation and understanding completely within the realm of physics forces us
to choose the indifference principle as more fundamental. This is simply because
the need for fine-tuning threatens to take us outside of where physics
or any of the other natural sciences can go.
Furthermore, as we have also seen, physics and the other sciences cannot delve
into the realm of ultimate explanation either. 

\section{Conclusion}

As we stressed in the conclusion of EKS, the introduction of the
 multiverse or ensemble idea is a fundamental change
in the nature of cosmology, because it aims to challenge one of
the most basic aspects of standard cosmology, namely the
uniqueness of the universe (see Ellis 1991, 1999 and references
therein). So far, research and discussion on such ensembles have
not precisely specified what is required to define them, although
some specific physical calculations have been given based on
restricted low-dimensional multiverses.

Our fundamental starting point has been the recognition that there
is an important distinction to be made between possible universes
and realised universes, and a central conclusion is that a really
existing ensemble or multiverse is not \textit{a priori} unique,
nor uniquely defined. It must somehow be selected for. We have
defined both the ensemble of possible universes
$\mathcal{M}$,\thinspace and ensembles of really existing
universes, which are envisioned as generated by a given primordial
process or action of an overarching cosmic principle, physical or
metaphysical. This effectively selects a really existing
multiverse from $\mathcal{M}$, and, as such, effectively defines a
distribution function over $\mathcal{M}$. Thus, there is a
definite causal connection, or \textquotedblleft law of laws,
\textquotedblright relating all the universes in each of those
multiverses. It is such a really existing ensemble of universes,
one of which is our own universe, \textit{not} the ensemble of all
possible universes, which provides the basis for anthropic
arguments. Anthropic universes lie in a small subset of
$\mathcal{M}$, whose characteristics we understand to some extent.
It is very likely that the simultaneous realisation of \emph{all}
the conditions for life will pick out only a very small sector of
the parameter space of all possibilities: anthropic universes are
fine-tuned in that sense. If cosmogonic processes or the operation
of a certain primordial principle selected and generated an
ensemble of really existing universes from $\mathcal{M}$, some of
which are anthropic, then, though we would require some
explanation for that process or principle, the fine-tuning of our
universe would not require any other scientific explanation. It
is, however, abundantly clear that \textquotedblleft really
existing ensembles\textquotedblright\ are \textit{not} unique, and
neither their properties nor their existence are directly
testable. Arguments for their existence would be much stronger if
the hypotheses employing them were fruitful in enabling new
investigations leading to new predictions and understandings which
are testable. However, so far this has not been the case. In our
view these questions - Issues 1 and 2 in this paper -- cannot be
answered scientifically with any adequacy because of the lack of
any possibility of verification of any proposed underlying theory.
They will of necessity have to be argued with a mixture of careful
philosophically informed science and scientifically informed
philosophy. And, even with this, as we have just seen, we seem to
fall short of providing satisfactory answers -- so far!

Another philosophical issue we have emphasized which has a strong
bearing on how we describe and delimit really existing multiverses
is that of realised infinity. From our careful discussion of this
concept, there is a compelling case for demanding that every
really existing ensemble contain only a finite number of universes
or universe domains.

There is strong support for both of two competing approaches --
that which honors the special character of our universe by
stressing the need for the fine-tuning of initial conditions and
the laws of nature, and that which locates its emergence in the
operation of primordial processes on a much more fundamental
generic or indifferent configuration. Both are undoubtedly at work
on different levels. The issues are:  which is more fundamental,
and whether the sciences themselves as they are presently
conceived and practiced can deal with ultimate fundamentals. Must
they yield that realm to metaphysics? Can metaphysics deal with
them? The relative untestability or unprovability of the
multiverse idea in the usual scientific sense is however
problematic -- the existence of the hypothesized ensemble remains
a matter of faith rather than of proof, unless it comes to enjoy
long-term fruitfulness and success. Furthermore in the end, \ the
multiverse hypothesis simply represents a regress of causation.
Ultimate questions remain: Why this multiverse with these
properties rather than others? What endows these with existence
and with this particular type of overall order? What are the
ultimate boundaries of possibility -- what makes something
possible, even though it may never be realised?\

As we now see, the concept of a multiverse raises many fascinating
issues that have not yet been adequately explored. The discussions
here should point and guide research in directions which will
yield further insight and understanding.

\section*{Acknowledgements}

\noindent We thank A. Aguirre, C. Harper, A. Lewis, A. Linde, A. Malcolm, and
J.P. Uzan for helpful comments and references related to this work, and two
anonymous referees for comments and suggestions which have enabled us to
correct errors and clarify our positions. GFRE and UK
acknowledge financial support from the University of Cape Town and the NRF
(South Africa).

\end{document}